\documentclass{aa}  
\usepackage{natbib,twoopt}
\usepackage[breaklinks=true]{hyperref} 
\bibpunct{(}{)}{;}{a}{}{,}             
\makeatletter
  \newcommandtwoopt{\citeads}[3][][]{\href{http://adsabs.harvard.edu/abs/#3}%
    {\def\hyper@linkstart##1##2{}%
     \let\hyper@linkend\@empty\citealp[#1][#2]{#3}}}
  \newcommandtwoopt{\citepads}[3][][]{\href{http://adsabs.harvard.edu/abs/#3}%
    {\def\hyper@linkstart##1##2{}%
     \let\hyper@linkend\@empty\citep[#1][#2]{#3}}}
  \newcommandtwoopt{\citetads}[3][][]{\href{http://adsabs.harvard.edu/abs/#3}%
    {\def\hyper@linkstart##1##2{}%
     \let\hyper@linkend\@empty\citet[#1][#2]{#3}}}
  \newcommandtwoopt{\citeyearads}[3][][]%
    {\href{http://adsabs.harvard.edu/abs/#3}
    {\def\hyper@linkstart##1##2{}%
     \let\hyper@linkend\@empty\citeyear[#1][#2]{#3}}}
\makeatother

\usepackage{graphicx}
\usepackage{txfonts}
\usepackage{subfigure}
\usepackage{longtable}
\usepackage{tabularx, blindtext}

\begin{document} 
 \title{VVVX-Gaia Discovery of a Low Luminosity Globular Cluster in the Milky Way Disk}
   \author{E.R. Garro \inst{1}
          \and
          D. Minniti\inst{1,2}
          \and
          M. G{\'{o}}mez\inst{1}
          \and
           J. Alonso-Garc{\'{i}}a \inst{3,4}
           \and
           R. H. Barb{\'{a}} \inst{5}
           \and
			B. Barbuy \inst{6}
			\and
			J. J. Clari{\'{a}} \inst{7}
			\and
			A.~N.~Chen{\'{e}}\inst{8}
			\and
			B.~Dias\inst{9}
			\and
			M.~Hempel \inst{1}
			\and
			V.~D.~Ivanov \inst{10}
			\and
			P.~W.~Lucas \inst{11}
			\and
			D.~Majaess\inst{12,13}
			\and
			F.~Mauro\inst{14}
			\and
			C.~Moni~Bidin \inst{14}
			\and
			T.~Palma\inst{7}
			\and
			J.~B.~Pullen\inst{1}
			\and
			R.~K.~Saito\inst{15}
			\and
			L.~Smith\inst{16}
			\and
			F.~Surot\inst{17}
			\and
		   S.~Ram{\'{i}}rez Alegr{\'{i}}a\inst{3}
		  \and
		   M.~Rejkuba\inst{11}
		  \and
			V.~Ripepi\inst{18}
          }
   \institute{Departamento de Ciencias F{\'{i}}sicas, Facultad de Ciencias Exactas, Universidad Andr{\'{e}}s Bello, Fern{\'{a}}ndez Concha 700, Las Condes, Santiago, Chile
   \and
 Vatican Observatory, Vatican City State, V-00120, Italy
 \and
 Centro de Astronom{\'{i}}a (CITEVA), Universidad de Antofagasta, Av. Angamos 601, Antofagasta, Chile
 \and
   Millennium Institute of Astrophysics, Santiago, Chile
   \and
   Departamento de Astronom{\'{i}}a, Av. Cisternas 1200 Norte, La Serena, Chile
   \and
      Universidade de S{\~a}o Paulo, IAG, Rua do Mat{\~a}o 1226, Cidade Universit{\'{a}}ria, S{\~a}o Paulo 05508-900, Brazil 
     \and
     Observatorio Astron{\'{o}}mico, Universidad Nacional de C{\'{o}}rdoba, Laprida 854, C{\'{o}}rdoba, Argentina
     \and
     Gemini Observatory, Northern Operations Center, 670 A'ohoku Place, Hilo, HI 96720, USA
     \and
     Instituto de Alta Investigaci{\'{o}}n, Universidad de Tarapac{\'{a}}, Casilla 7D, Arica, Chile 
     \and
     European Southern Observatory, Karl-Schwarszchild-Strasse 2, 85748 Garching bei Muenchen, Germany
     \and
     Department of Astronomy, University of Hertfordshire, Hertfordshire, UK 
     \and
Mount Saint Vincent University, Halifax, Nova Scotia, Canada
\and
Saint Mary's University, Halifax, Nova Scotia, Canada
\and
Instituto de Astronom{\'{i}}a, Universidad Cat{\'{o}}lica del Norte, Av. Angamos 0610, Antofagasta, Chile
	\and
Departamento de F{\'{i}}sica, Universidade Federal de Santa Catarina, Trinidade 88040-900 Florian{\'{o}}polis, SC, Brazil 
\and
Institute of Astronomy, University of Cambridge, Madingley Road, Cambridge, CB3 0HA, UK
\and
Instituto de Astrof{\'{i}}sica de Canarias, V{\'{i}}a L{\'{a}}ctea S/N, E-38200, La Laguna Tenerife, Spain 
\and
INAF-Osservatorio Astronomico di Capodimonte, Salita Moiariello 16, 80131, Naples, Italy
 }
  \date{Received 21 August 2020; Accepted 29 September 2020}

  \abstract
   {
Milky Way globular clusters (GCs) are difficult to identify at low Galactic latitudes because of high differential extinction and heavy star crowding. 
The new deep near-IR images and photometry from the VISTA Variables in the Via L{\'{a}}ctea Extended Survey (VVVX) allow us to chart previously unexplored regions.}
   {
   Our long term aim is to complete the census of Milky Way GCs. The immediate goals are to estimate the astrophysical parameters for the newly discovered globular cluster candidates, measuring their reddenings, extinctions, distances, total luminosities, proper motions, sizes, metallicities and ages.
   }
   {
   We use the near-IR VVVX survey database, in combination with Gaia Data Release 2 (DR2) optical photometry and proper motions (PMs), and with the Two Micron All Sky Survey (2MASS) photometry, to search and characterise new GCs within the Southern Galactic plane ($|b|<5^{\circ}$). 
   }
   {
We report the detection of a heretofore unknown Galactic Globular Cluster at $RA =$ 14:09:00.0; $DEC=-$65:37:12 (J2000) corresponding to  $l = 310.828$ deg; $b = -3.944$ deg in galactic coordinates.  We calculate a reddening of $E(J-K_{s})=(0.3\pm 0.03)$ mag and an extinction of $A_{K_s}=(0.15\pm 0.01)$ mag for this new GC. Its distance modulus and corresponding distance were measured as $(m-M)=(15.93\pm 0.03)$ mag and $D=(15.5\pm 1.0)$ kpc, respectively. We also estimate the metallicity and age by comparison with known globular clusters and by fitting PARSEC and Dartmouth isochrones, finding $[Fe/H]=(-0.70\pm 0.2)$ dex and $t=(11.0\pm 1.0)$ Gyr.  The mean GC PMs from Gaia DR2 are $\mu_{\alpha^\ast}=(-4.68 \pm 0.47 )$ mas $yr^{-1}$ and $\mu_{\delta}=(-1.34 \pm 0.45)$ mas $yr^{-1}$.
The total luminosity of our cluster is estimated to be $M_{Ks}=(-7.76\pm 0.5)$ mag. The core and tidal radii from the radial density profile are $r_c\sim 2.1'$ (4.6 pc) and $r_t=6.5'$ (14.6 pc) at the cluster distance. 
}
{
We have found a new low-luminosity, old and metal-rich globular cluster, situated in the far side of the Galactic disk, at $R_{G}=11.2$ kpc from the Galactic centre, and at $z=1.0$ kpc below the plane. 
Interestingly, the location, metallicity and age of this globular cluster are coincident with the Monoceros Ring (MRi) structure.}
  
   
   \maketitle
\section{Introduction}
While most Galactic globular clusters (GCs) have been found using optical wavelength observations, large infrared surveys have made possible to discover many more of them in recent decades. The main surveys are the Two Micron All Sky Survey (2MASS) observed in both hemispheres with two 2-m telescopes, one located at the Cerro Tololo Inter-American Observatory and the other at Kitt Peak National Observatory \citep{2006AJ....131.1163S,2003yCat.2246....0C}, the Galactic Legacy Infrared Midplane Survey Extraordinaire (GLIMPSE) based on observations with the NASA/ESA Spitzer Space Telescope \citep{2005ApJ...630L.149B}, the UKIDSS Galactic Plane Survey (GPS) based on observations at the UKIRT 4-m telescope in Hawaii \citep{Lucas_2008}, the NASA Wide Infrared Satellite Experiment (WISE -- \citealt{2010AJ....140.1868W}), and the VISTA Variables in the Via L{\'{a}}ctea  Survey based on observations at the VISTA 4-m telescope (VVV -- \citealt{2010NewA...15..433M, 2012A&A...537A.107S}). This last survey, in particular, was expanded, becoming the VISTA Variables in the Via L{\'{a}}ctea eXtended Survey (VVVX) with observations from years 2017 to 2021, also using the Visual-Infrared Camera (VIRCAM) at the VISTA telescope at ESO Paranal Observatory. The VVV has already enabled discovery and characterization of numerous new clusters (e.g. 
 \citealt{2011A&A...527A..81M, 2011A&A...532A.131B, 2011A&A...535A..33M, Minniti_2017,2017A&A...600A.112I,2019A&A...628A..45G,10.1093/mnrasl/slz010}), and its extension VVVX is delivering on the promise to find even more in the extended area (e.g. \citealt{2018MNRAS.481.3902B}). \\
However, the search for new GCs is still an intricate undertaking, since many considerations about the formation and evolution of GCs have to be included in this type of study. Clusters are not likely to contribute much to the stellar halo (e.g. \citealt{2011A&A...534A.136M,2019A&A...625A..75K,2020MNRAS.493.3422R}), but it is known that the clusters disrupt, especially those with orbit going through the bulge and disk (e.g. \citealt{2003MNRAS.340..227B, Kruijssen_2011}). Hence in the dense disk/bulge areas, there may be many small clusters that have lost part of their mass or had fewer stars from the beginning given the cluster initial mass function that predicts many more low mass clusters forming than are observed today among old GCs. Therefore, the detection of some GCs may be complicated due to their low-luminosities and also the presence of dust and field stars that probably cover these objects.

\section{Discovery of Garro01}
\begin{figure}[h]
\centering
\includegraphics[width=18cm, height=6cm]{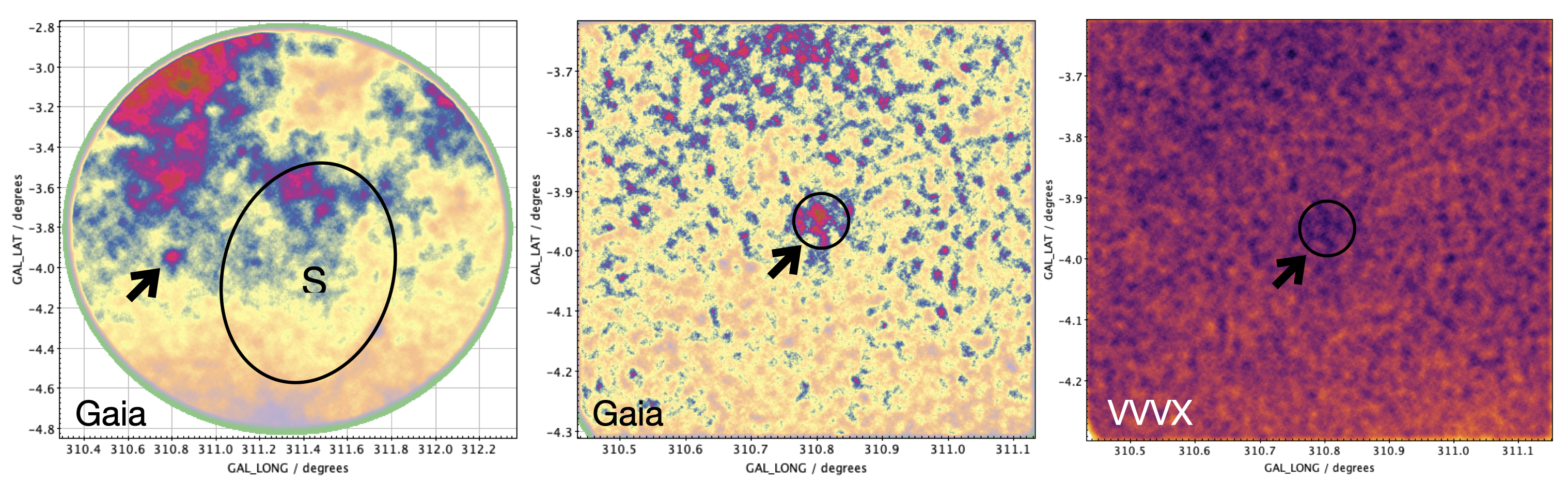} 
\caption{Left panel: Gaia stellar density map for a $r=2$ degree field in Galactic coordinates, indicating the new GC with an arrow, along with the size of the HI emission of the Circinus galaxy \citep{1977A&A....55..445F}. Middle and right panel: Density maps of the zoomed $42' \times 42'$ region around Garro01 using optical Gaia data and VVVX near-IR data, respectively, where the circle indicates the approximate cluster size. Note the redder areas are representative of overdensities while the yellower areas are lower densities.}
\label{density}
\end{figure}

The search and physical characterization of Galactic GCs is one of the scientific goals of our VVVX survey, already yielding some results (\citealt{2018MNRAS.481.3902B, Barb__2019}, Borissova et al. 2020, Minniti et al. 2020, Garro et al. 2020 in prep., Obasi et al. 2020 in prep.). Here, we report the discovery and confirmation of VVVX-GC-140900-653712 (hereafter Garro01 for short), a distant new GC deeply embedded in the Milky Way (MW) disk.
Following the same procedure described by \cite{Minniti_2017}, this cluster was found in the VVVX database as a clear overdensity of red giant stars above the background.
This GC lies in a complex low-latitude region, with variable extinction, high stellar density, containing evidence for star formation towards the Galactic plane, and also close (at $\sim$1${^\circ}$) in projection to the Circinus Seyfert 2 galaxy that is very extended on the sky \citep{1977A&A....55..445F}.
From the first visual analysis for example, Garro01 looks like a low-luminosity GC compared with our previous finding of VVV$-$GC$-05$, which is a metal-poor $[Fe/H]=-1.3$ dex GC located in the same region of the sky, at $l=330{^\circ}$ , $b=-1.8{^\circ}$, and $D=7.5$ kpc \citep{Minniti_2017, 2018ApJ...863...79C}.
We also observe Garro01 as a clear excess of stars above the background in the optical, where the Gaia Data Release 2 (DR2) \citep{2018A&A...616A...1G} source density map of the region (Figure \ref{density}) clearly shows a nearly circular region. We have visually selected it like a circle with $r\sim 2.5'$ centred at equatorial coordinates $RA=$ 14:09:00.0; $DEC= -$65:37:12 (J2000), corresponding to Galactic coordinates $l= 310.8278{^\circ}$, $b= -3.9442{^\circ}$. 

\section{Physical Characterization of the New Globular Cluster Garro01}
We used a combination of near-IR and optical data, obtained with the VVVX, the 2MASS and Gaia DR2 surveys, in order to measure the main parameters for Garro01. While for the 2MASS and Gaia~DR2 we downloaded the available data from their respective repositories, for the VVVX we use a preliminary version of the point spread function (PSF) photometry that we are developing for its whole footprint based on \cite{2018A&A...619A...4A} VVV PSF photometry. This  VVVX PSF photometry was already used by \cite{Barb__2019} for the discovery and characterization of FSR-1758, a giant metal-poor retrograde GC \citep{2019ApJ...882..174V}, and by \cite{2018A&A...616A..26M} for the discovery of extra tidal RR Lyrae stars in the metal-poor bulge GC NGC~6266 (M62). For our analysis, we put the VVVX PSF photometry of our region of interest in the 2MASS $JHK_s$ photometric system. \\

\begin{figure}[h]
\centering
\includegraphics[width=12cm, height=5.5cm]{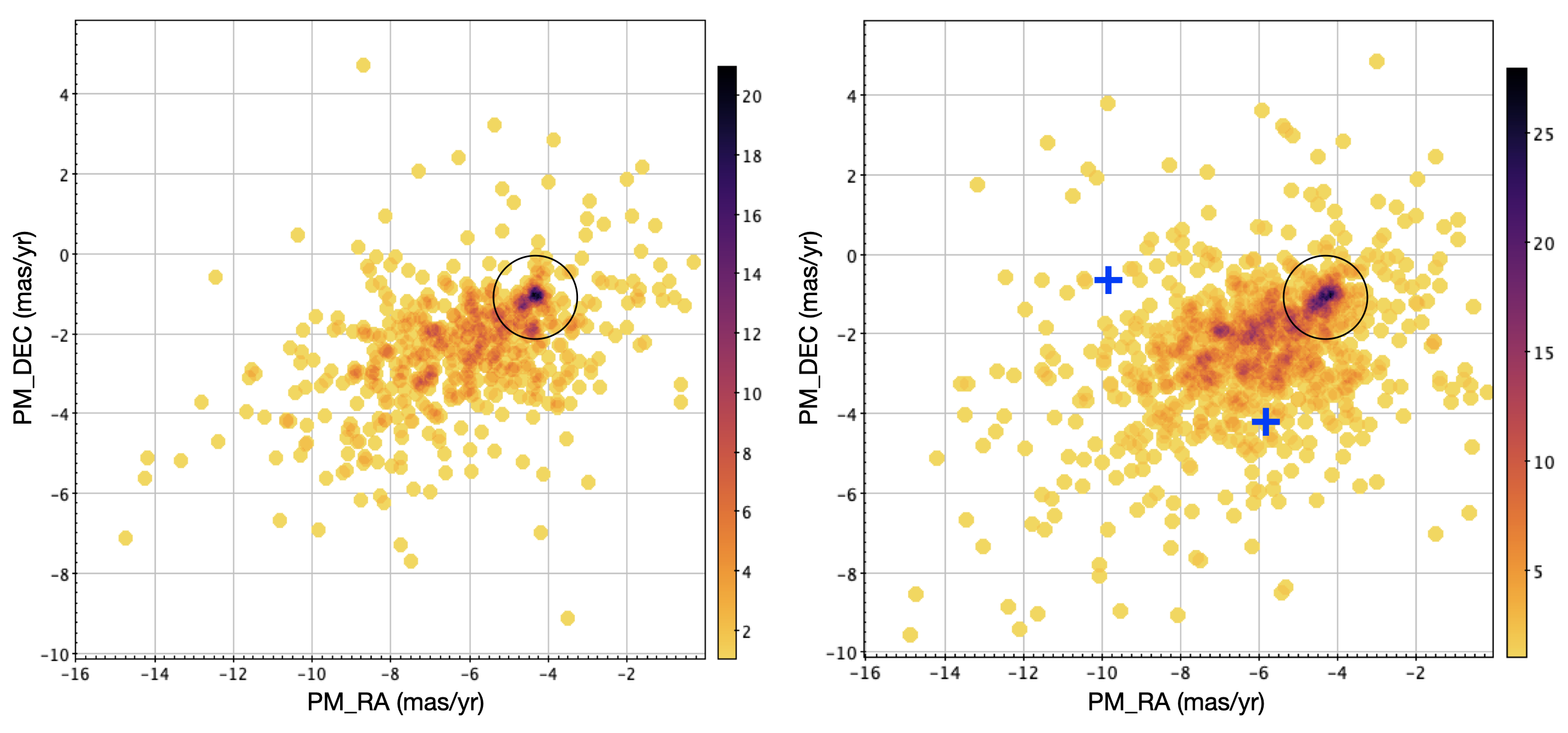} 
\caption{Vector PM diagrams for the bright sources ($Ks<15$ mag) matched in the 2MASS+Gaia catalogues (right panel), and the fainter sources ($Ks>13$ mag) matched in the VVVX+Gaia catalogues (left panel). The black circle indicates the cluster selection, while blue crosses the position of the two RR Lyrae found within $10'$ of the cluster centre. The colour bars indicate a higher (towards black colour) and lower (towards yellow colour) concentration.}
\label{vpmd}
\end{figure}

\begin{figure}[h]
\centering
\includegraphics[width=10cm, height=10cm]{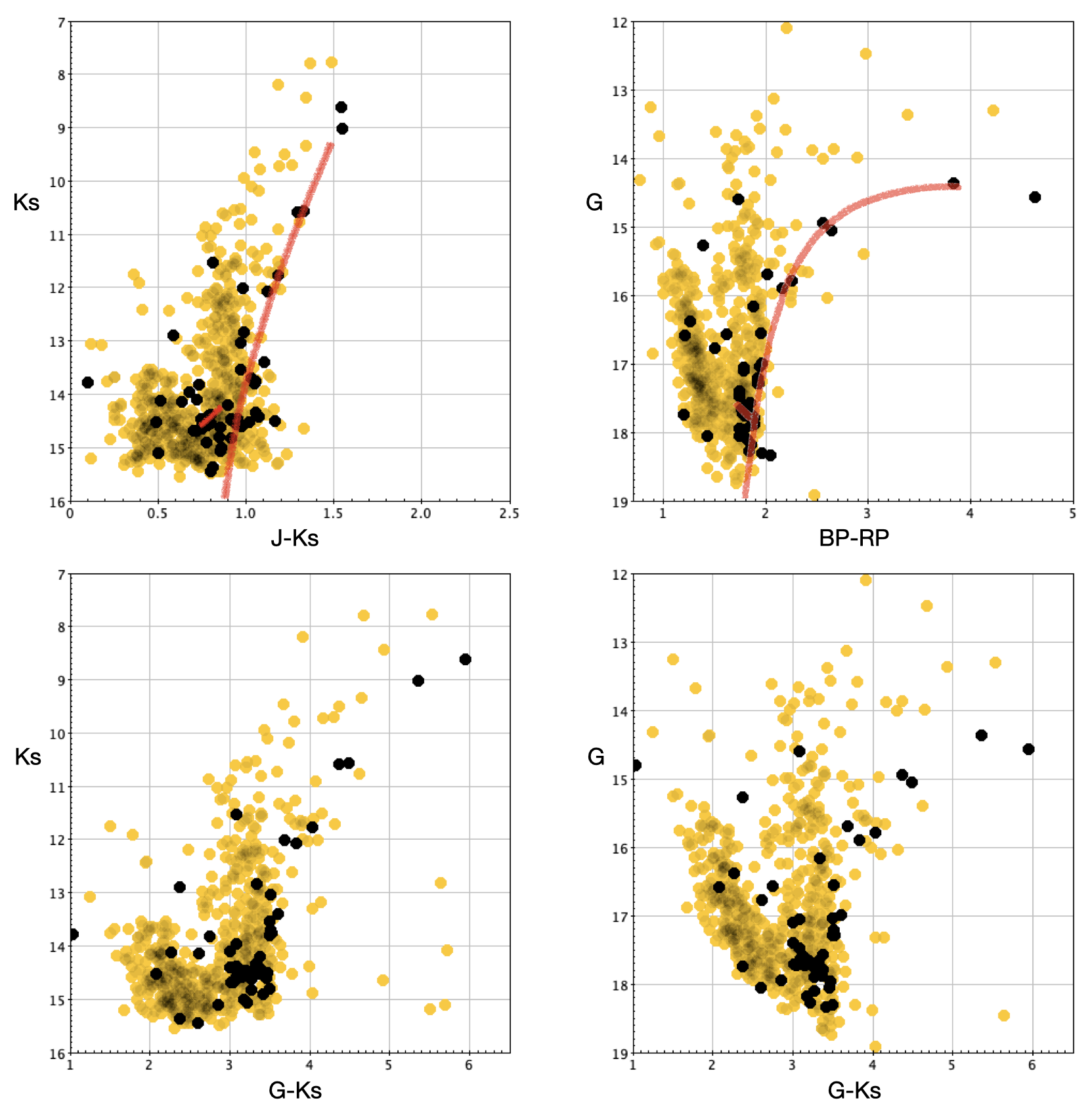} 
\caption{Optical and near-IR CMDs for the GC Garro01 field from 2MASS (black points) and Gaia or VVVX (yellow points), showing the PM-selected cluster members. Note the well-defined cluster RGB located lower and to the red side of the diagrams, indicating that the fiducial RGB from the GC 47 Tuc from \citet{Cohen_2015} and \citet{Babusiaux_2018} is overplotted in the top panels, appropriately shifted to the cluster distance and reddening this is a distant and reddened cluster.}
\label{cmd}
\end{figure}

We build-up a clean, decontaminated catalogue of highly-probable cluster members, benefiting from the precise astrometry and proper motions (PMs) from Gaia DR2, but also matching 2MASS+Gaia and VVVX+Gaia catalogues as a means to include both brighter (since stars with $Ks<11$ mag are saturated in VVVX photometry) and fainter sources. We first discard all nearby stars with parallax $>0.5$ mas. Then, we inspect the vector PM (VPM) diagram (Figure \ref{vpmd}), that shows a sharp peak on top of the broad stellar background distribution that we interpret as the cluster mean PM. These mean PMs as measured by Gaia~DR2 are: $\mu_{\alpha^\ast} = (-4.68 \pm 0.47)$  mas yr$^{-1}$; $\mu_{\delta} = (-1.35 \pm 0.45)$ mas yr$^{-1}$. Therefore, we select as probable cluster members all stars within 1 mas yr$^{-1}$ from the mean cluster PMs. Figure \ref{vpmd} shows the VPM diagrams for sources with $Ks<15$ mag matched in the 2MASS+Gaia catalogues, and sources with $Ks>13$ mag matched in the VVVX + Gaia catalogues, respectively. In the same Figure, the stars located outside 1 mas yr$^{-1}$ circle selected on the VPM diagram, but within the selected radius of $2.5'$ of the cluster centre are shown in yellow. These stars are predominantly coming from the surrounding field that includes disk foreground/background stars.
 \\
The Gaia+2MASS colour-magnitude diagrams (CMDs) are shown in Figure \ref{cmd}, highlighting the tight red giant branch (RGB) of the PM selected cluster members.
The fact that the cluster RGB is preferentially fainter and redder than the field stars indicates that the field stars belong mostly to a foreground less-reddened population in the Galactic plane. 
This rules out a window in the dust distribution, as such windows are also detected like stellar overdensities (e.g. \citealt{2018ASSP...51...63M,10.1093/mnrasl/slaa028}). 
For comparison, the fiducial RGB from the GC 47\,Tuc (taken from \citealt{Babusiaux_2018} and \citealt{Cohen_2015}) is overplotted in red in the Gaia $G$ vs $(BP-RP)$ and in the VVVX $Ks$ vs $(J-Ks)$, respectively, applied for all other colour indexes (moving it of $\Delta G=3.8$ mag and $\Delta (BP-RP)=0.39$ mag), showing excellent agreements with the RGB of Garro01. We bring attention to the overdensity of stars around $G\sim17.70$ in the Gaia CMD, which is well fitted with the slanted line corresponding to the location of the Red Clump (RC) core He burning stars in 47 Tuc. \\

\begin{figure}[h]
\centering
\includegraphics[width=8cm, height=8cm]{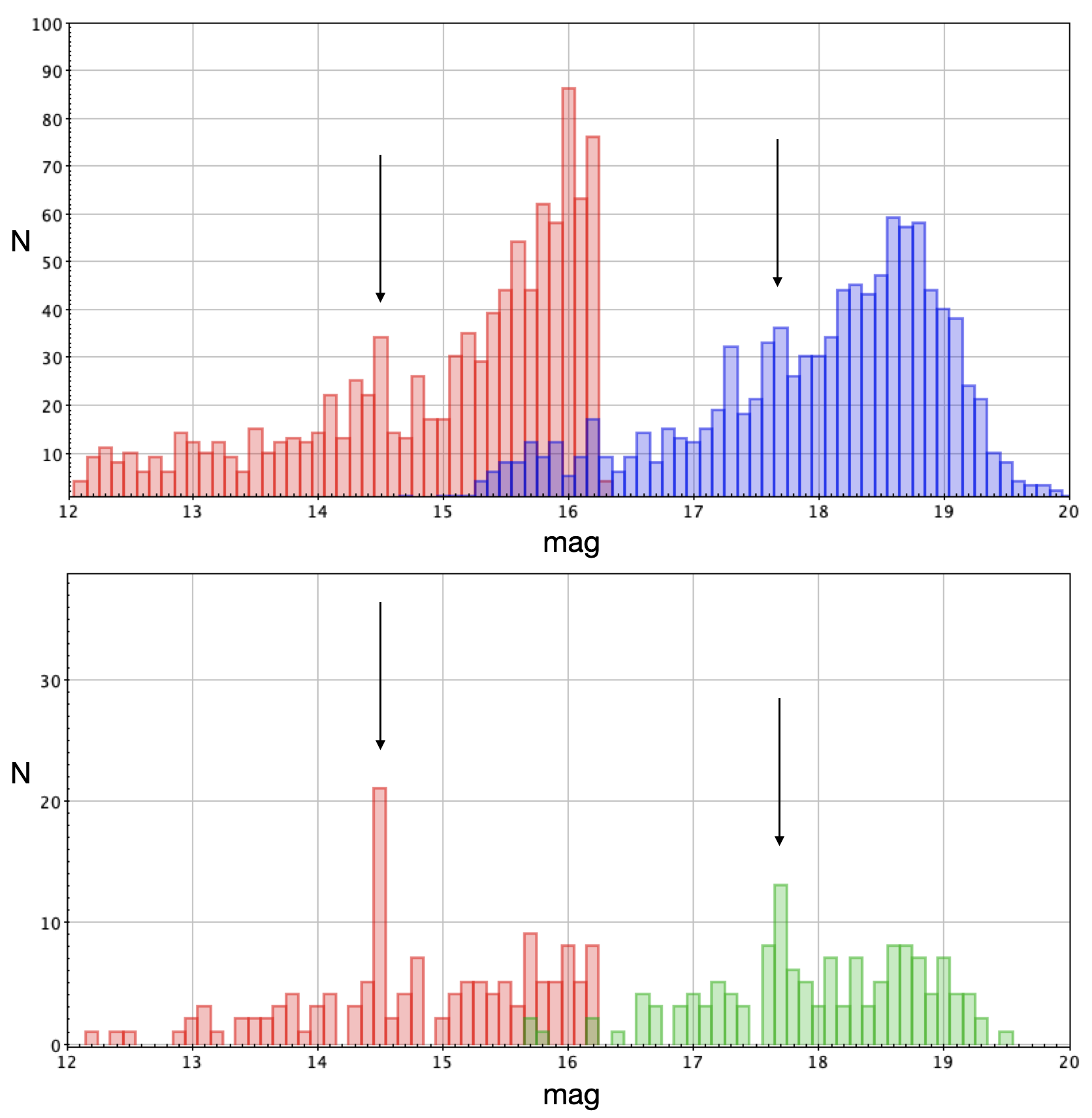}
\caption{Top panel: Luminosity functions for the GC Garro01, showing the VVVX Ks-band photometry in red on the left, and the Gaia DR2 G-band photometry in blue on the right. The arrows point  the location of the RC. 
Bottom panel: The same diagram showing only the counts of PM selected sources.}
\label{lfs}
\end{figure}

Figure \ref{lfs} shows the G and Ks bands luminosity functions for a $3'$ radius field centred on the cluster, for the all red giant stars with $J-Ks > 0.7$ mag (top panel), as well as for the PM selected red giants (bottom panel). These luminosity functions exhibit clearly the peaks due to the cluster RC giants at $G_{RC}=(17.70 \pm 0.05)$ mag, and $Ks_{RC}=(14.48 \pm 0.05)$ mag, respectively. \\
 
\begin{figure}[h]
\centering
\includegraphics[width=10cm, height=10cm]{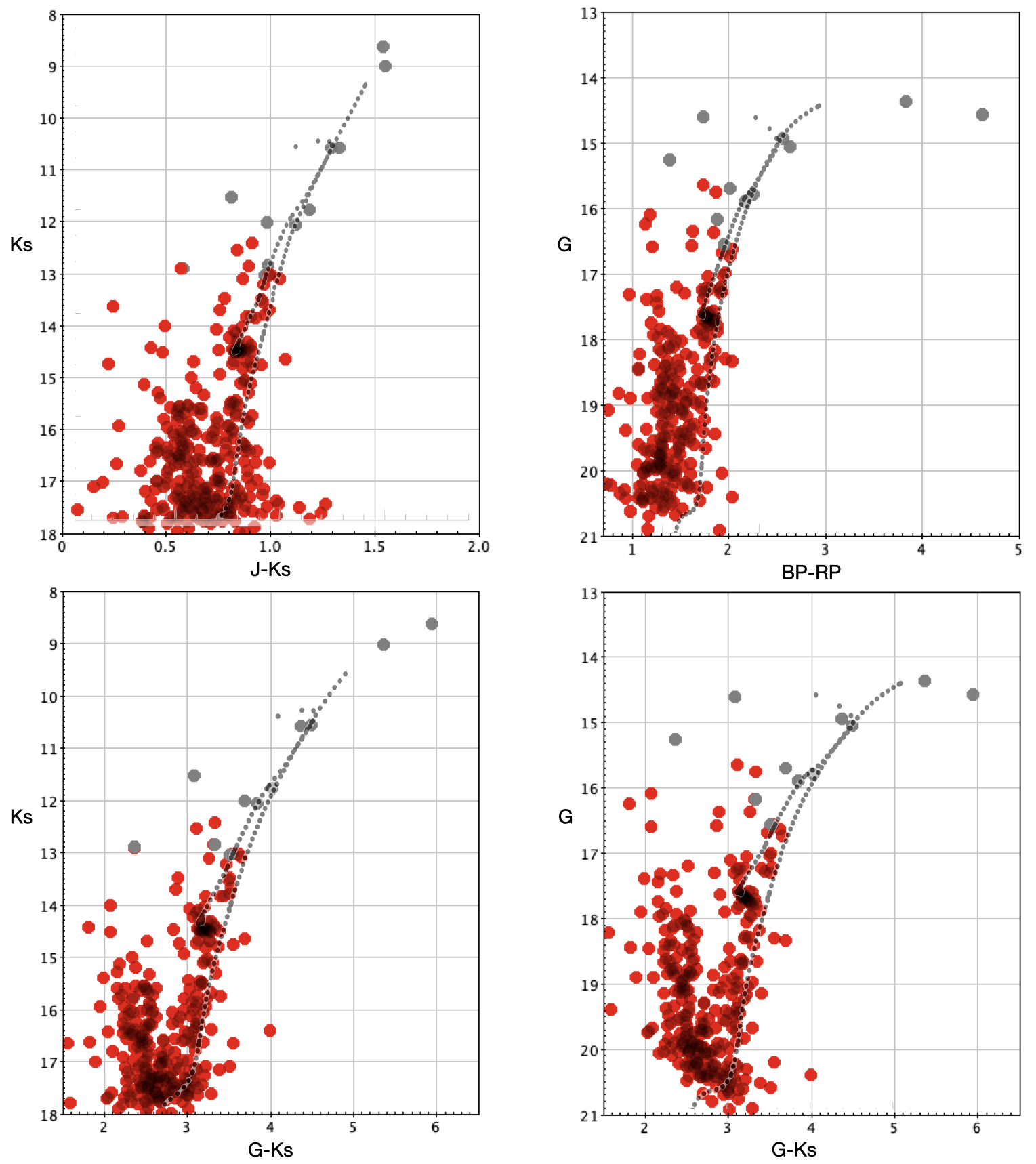} 
\caption{PM decontaminated CMDs for the GC Garro01, showing the VVVX near-IR CMD (top left), the Gaia DR2 optical CMD (top right), and the Gaia-VVVX optical-IR CMDs (bottom panels). Red points are VVVX-Gaia matched sources, while grey points are 2MASS-Gaia matched sources. The black dotted line shows the fit to a PARSEC isochrone with metallicity $[Fe/H]=-0.70$ dex and age $t=11.0$ Gyr.
Note the well defined cluster RGB and RC on all of the CMDs, and also residual field contamination on the blue side arising from a fraction of Galactic foreground field stars with similar PMs as the GC.}
\label{cmdiso}
\end{figure}

The resulting PM-cleaned CMDs from the Gaia+2MASS+VVVX data are displayed in Figure \ref{cmdiso}. 
These CMDs show that the RGB is narrow and well defined, and the  cluster RC is clearly seen. 
Note also the presence of blue star residual contamination. 
These stars are too bright to be blue stragglers belonging to Garro01, and we argue that this contamination arises from a fraction of Galactic foreground field stars with similar PMs as the GC.\\
We estimate the reddening and extinction in this field following the maps of \cite{2018A&A...609A.116R} in the near-IR and \cite{2011ApJ...737..103S} in the optical, and taking advantages from the mean magnitude of RC giants. In detail, we assume the RC stars calibration of \cite{2018A&A...609A.116R}, where the absolute magnitude in Ks-band is $M_{Ks}=(-1.601 \pm 0.009)$ mag and the relative colour is $(J-Ks)_0= (0.66 \pm 0.02)$ mag, obtaining $E(J-K_{s}) = (0.30\pm 0.03)$ mag and $A_{Ks}=(0.15\pm 0.01)$ mag, respectively. In particular,  we noted that within $\sim 10'$  from the GC centre the reddening seems to be fairly uniform, with variations $\Delta E(J-K_{s})<0.05$ mag. \\
The GC distance modulus can then be measured, adopting $A_{Ks}/E(J-Ks)=0.5$ mag \citep{2018A&A...616A..26M}, yielding $(m-M)_{0}=(15.93\pm0.03)$ mag for Garro01, and therefore the heliocentric distance is $D=(15.5\pm1.0)$ kpc, placing this GC at $R_G=11.2$ kpc from the Galactic centre \citep{2016ARA&A..54..529B}, assuming $R_{0}=8.3$ kpc \citep{2013ApJ...776L..19D}. 
We would like to point out that we also compared our results with other K-band calibration methods, in order to achieve more robust parameters. We used RC magnitude theoretical calibrations from \cite{2002MNRAS.337..332S} and \citet{2002ApJ...573L..51A}, finding a perfect agreement ($(m-M)_{0}=15.87$ mag, $D=15.0$ kpc, and $(m-M)_{0}=15.93$ mag, $D=15.3$ kpc, respectively).
Simultaneously, we derive the parameters using Gaia DR2 photometry. Adopting the extinction $A_{Ks}=(0.15\pm 0.01)$ mag, we obtain the equivalent value in G-band of $A_{G}=(1.2\pm 0.2)$ mag \citep[employing the V-band extinction value
$A_{V} = 1.4$ mag from][]{2011ApJ...737..103S} and so the reddening of $E(BP-RP)=(0.39\pm 0.06)$ mag. These values are used to measure the distance modulus of $(m-M)_{0}=(16.0\pm0.2)$ mag and heliocentric distance $D=(15.9\pm 1.0)$ kpc, in excellent agreement with the 2MASS+VVVX photometry.
The distance measurement of $D=15.5$ kpc yields a scale of $1'= 2.25$ pc, placing this cluster at a height of about 1.0 kpc below the Galactic plane. \\

Further, we carried out the radial density profile of Garro01, in order to determine its size. First of all, we checked the central position, because of an offset in its position may alter the radial density profile. Indeed, we found that the new centre is $RA=212.23$ deg and $DEC= -65.628$ deg (shifted to $\Delta RA=2.1'$ and $\Delta DEC=0.48'$ from the initial coordinates). Using the resulting centre, we divided our sample into six radial bins (out to a radius of $1.5'$) and we calculated the area in each bin. After that, we derive the density inside the circular annuli as the total number of stars over the area. Finally, we overplotted the King profile \citep{1962AJ.....67..471K} that well-reproduces density points, obtaining a core radius of $r_c=(2.1\pm 1.5)$ arcmin is equivalent to a physical size $r_c=(4.6\pm 3.1)$ pc, and a tidal radius of $r_t=6.5^{+11}_{-1.9}$ arcmin, corresponding to $15^{+25}_{-4}$~pc, consistent with the typical galactic GC sizes as listed in the 2010 \cite{1996AJ....112.1487H} compilation. We note that the Poisson errors are very large, due to a poor statistics, and the background level was supposed to be zero since the catalogue is assumed clean from contaminants.\\

The integrated GC absolute magnitude is estimated coadding the RGB stars from the PM decontaminated diagrams and employing the GC size of 6.5 arcmin. We find $M_{Ks} = (-7.76\pm 0.5)$ mag, equivalent to $M_{V}=( -5.26\pm 1.0)$ mag for typical GC mean colours $V-K= (2.5\pm 0.94)$ mag. Certainly, this represents a lower limit since sub-giant branch and main-sequence stars are below the observational limit. Indeed, benefiting from strong resemblance 47\,Tuc, we also estimated the total luminosity for Garro01. Briefly, we calculated the integrated absolute magnitude, coadding the RGB 47\,Tuc stars, 
and scaling to the value $M_V=-9.42$ mag from 2010 \cite{1996AJ....112.1487H} catalogue. 
This results in an absolute magnitude $M_V=-5.62$ mag for Garro01, placing this GC on the low-luminosity tail of the MW GCLF, $\sim 2$ magnitudes fainter than the peak of the MW GCLF \citep[$M_{V}=(-7.4\pm 0.2)$ mag from][]{Harris_1991,ashman_1998}.\\

Finally, the cluster metallicity is derived following two different methods. Firstly, we compare Garro01 with the fiducial GC 47 Tuc (age $t=11.8$ Gyr, $[Fe/H]=-0.72$ dex, $[\alpha/Fe]=+0.4$ dex, $D= 4.5$ kpc from \citealt{10.1093/mnras/stx378}), making reddening adjustments in order that evolutionary sequences of both GCs coincide. Secondly, we use the fitting-isochrone method, preferring both PARSEC \citep{Marigo_2017} and Dartmouth \citep{2008ApJS..178...89D} isochrones. In order to have a good fit of the stellar isochrones in the near-IR and optical CMDs (Figure \ref{cmdiso}), we adopt the resulting values of reddening, extinction and distance modulus calculated from the VVVX and Gaia photometries. Both methods provide a metallicity of $[Fe/H]=(-0.70 \pm 0.1)$ dex and $\alpha$-enhanced between 0 and $+0.4$ dex . However, the cluster appears to be slightly younger than 47 Tuc because we estimate an age $t=(11 \pm 1)$ Gyr (from stellar isochrones). Briefly, the best fitting age and metallicity is obtained by iteratively fitting isochrones by comparing our data with isochrones generated with different ages and metallicities and selecting the best by-eye fit. First we fix the age and vary metallicity. Then we fix the metallicity and search for the best fitting age. In this way, we also deduce an estimation of the metallicity and age errors, until the fitting-isochrones do not reproduce all evolutionary sequences in both optical and near-IR CMDs. It is important to note that this is a rough estimation of the age since the main-sequence turn-off is below to the magnitude lower limit.\\

In addition to the comparison with the GC 47\,Tuc, it is appropriate to make some comparisons with other known GCs. For this purpose we considered Lynga 7 ($[Fe/H]= -1.01$ dex, $D=8.0$ kpc, \citealt{2016PASA...33...28B}), and NGC 5927 ($[Fe/H]=-0.32$ dex, $D=8.2$ kpc, \citealt{2020MNRAS.491.3251P}), which are located in the same region of the Galactic plane, and also the 9 Gyr old open cluster NGC 6791 ($[Fe/H] = +0.3$ dex, $D= 4.1$ kpc, \citealt{Boesgaard_2009}). These comparisons do not match as well as 47 Tuc, exhibiting in particular obvious differences in the colour of the RGB and also the position of the RC. \\

We also search for more evidence of an old GC population such as the RR Lyrae stars, finding two RR Lyrae candidates from the Gaia DR2 catalogue \citep{Clementini_2019} within $10'$ of this GC. These are Gaia DR2 5851208911444645504, located $8.5$ arcmin away from the GC centre at $RA=$ 14:10:16.06 and $DEC= -$65:40:28.6 (J2000), with mean magnitude $G=16.94$ mag, amplitude $A= 0.73$ mag, and period $P= 0.570059$ days; and Gaia DR2 5851209426894334592, located $8.6$ arcmin away at $RA=$ 14:10:23.49 and $DEC= -$65:36:56.0 (J2000), with mean magnitude $G=16.33$ mag, amplitude $A= 0.94$ mag, and period $P= 0.595121$ days. Both RR Lyrae are compatible with fundamental mode pulsators (RRab type). Their distances are measured using the period-luminosity relation from \cite{Clementini_2019}. We believe that only Gaia DR2 5851208911444645504 may be a real cluster member, with a heliocentric distance of $D_{RRL}=15.9$ kpc, whereas for Gaia DR2 5851209426894334592 we calculate a much shorter distance $D_{RRL}=12.4$ kpc. However, for both stars their Gaia DR2 PMs ($\mu_{\alpha^\ast}=(-9.685 \pm 0.082)$  mas yr$^{-1}$; $\mu_{\delta}=(-0.603 \pm 0.102)$  mas yr$^{-1}$; and $\mu_{\alpha^\ast}=(-5.866 \pm 0.085)$  mas yr$^{-1}$; $\mu_{\delta}=(-4.190 \pm 0.098)$ mas yr$^{-1}$, respectively) are  $>4 \sigma$ different from the mean GC PM. This result calls into question the cluster membership for both variable stars, therefore, until the RR Lyrae membership is better established with additional data, we prefer to be cautious and adopt the distance measured using the RC giants, $D=15.5$ kpc. 

The physical parameters of the new GC Garro01 are summarised in Table \ref{table1}, along with their respective uncertainties ($1 \sigma$).

\section{Discussion}
The discovery of this new GC demonstrates that the Galactic GC census is incomplete, so many more objects like this may still be found at low Galactic latitudes, hidden in very crowded, heavily reddened regions (e.g. \citealt{2000A&A...362L...1I,2000AJ....119.2274I,2005A&A...442..195I,2007A&A...474..121B,2008A&A...489..583K,2011MNRAS.416..465L,2015MNRAS.446..730V}). The known low-luminosity GCs are very rare and located at high Galactic latitudes \citep{Van_Den_Bergh_2003}, and  the low-latitude ones would be particularly difficult to find. 

In addition, we have to consider that the history of GCs is complex, as well as that of our Galaxy. It is known that the MW has suffered past merging events (e.g. \citealt{1994Natur.370..194I,1999Natur.402...53H, Helmi2018,10.1093/mnras/sty982, Myeong_2018}). Further, photometric detections of stellar streams and substructures, especially in the Galactic halo \citep{Bell_2008}, are consistent with predictions from cosmological simulations that link stellar streams to accreted dwarf satellite galaxies \citep{Bullock_2005}. As a consequence, many GCs may be associated with these merging events (e.g. \citealt{2019A&A...630L...4M, 10.1093/mnras/stz171}).\\
A known structure close to the position of Garro01 is the Monoceros Ring (MRi, a.k.a Galactic anti-centre stellar structure, GASS). It was found by \cite{2002ApJ...569..245N} and could represent a fitting example of a past merging event. However, the nature of MRi is still a topic of discussion. \citet{2005MNRAS.362..906M}, \citet{2005ApJ...626..128P}, \citet{Morganson_2016}, \citet{10.1093/mnras/stx3048} modelled the remnant of a tidally-disrupted satellite galaxy, the Canis Major dwarf galaxy \citep{2003MNRAS.340L..21I,2003ApJ...588..824Y, 2003ApJ...594L.119C, 2003ApJ...594L.115R, Frinchaboy_2006,2006ApJ...651L..29G, Conn_2008,2011ApJ...726...47S}. These simulations reveal that the MRi feature should be present in the field studied here, as shown for example by
\citet[][see their Figs. 12 and 13]{Conn_2008}. Also, spectroscopic studies revealed that this feature is moderately metal-poor, with mean $[Fe/H] = -0.8$ dex \citep{Li_2012}, which is in agreement with the result presented here: $[Fe/H]=(-0.7 \pm 0.1)$ dex. However, an alternative scenario (e.g. \cite{2006A&A...451..515M,2011A&A...527A...6H,Kalberla_2014, Sheffield_2018}) argued that the MRi is merely the Galactic warp and flare, composed by stars from the MW disk kicked out to their current location due to interactions between a satellite galaxy and the disk. \\ 
These results are controversial, revealing the complexity of the regions under study. This fact makes it all the more valuable to have star clusters that may be associated with the MRi structure. Indeed, \cite{Frinchaboy_2006} searched for open and GCs associated with the MRi structure, concluding that Berkeley 29, Saurer 1, Tombaugh 2, Arp-Madore 2, Palomar 1, NGC 2808, and NGC 5286 were possibly associated, while BH176 is not. We note that the classification of BH 176 as a globular cluster has been recently questioned by \citet{2014A&A...570A..48S} and \citet{2018A&A...619A..13V}, and it might be a unique case. On the other hand, other studies \citep{2007AJ....133.1058C,2006MNRAS.368L..77M} concluded that old open clusters are not consistent with the models for the MRi, that is probably just related to the Galactic warp and flare. Using more recent photometric and spectroscopic observations,  \cite{2014AAS...22344212F} argued that Berkeley 29 ($[Fe/H]=-0.44$ dex, $D=13.4$ kpc, $R_{G}=21.1$ kpc) and Saurer 1 ($[Fe/ H]=-0.38$ dex, $D=13.1$ kpc, $R_{G}=20.2$ kpc) are associated with MRi, while again BH176 is not.\\

Our main goal is to understand the real nature of Garro01 and derive all parameters that describe it. For this reason, we have searched for possible association to other GCs in that region: Lynga 7, NGC 5927, 47 Tuc, BH176, and others. As mentioned in Section 2, we found very close similarity with 47\,Tuc, so we conclude that Garro01 is a new Galactic GC. Also, given the metallicity ($[Fe/H]=-0.7$ dex) and the location ($D=15.5$ kpc, $R_G=11.2$ kpc) of Garro01 in close proximity of the MRi structure, it is attractive to consider its possible association with MRi. \\
The field star contamination, in the same field where Garro01 is located, overwhelms the GC stars, but the vast majority of the field stars along this line of sight seem to be brighter and less reddened and therefore located in the foreground (Figure \ref{cmd}). They also have PMs whose mean values are different from those of the new Garro01 (Figure \ref{vpmd}). Nonetheless, it is crucial to obtain radial velocities (RVs) for this cluster in order to constrain its orbital properties and to definitely confirm its association with the MRi structure. Also, we note that, according to the maps of \cite{2006A&A...451..515M}, there should be no warp in the specific direction of the field studied here at Galactic longitude $l=311$ deg. 
\\
Unfortunately, reliable Gaia DR2 RVs are not available for this cluster. A thorough search revealed that there is an excess of stars with $RV>200\ km/s$ in this field, although all these sources are too bright to be cluster members. Although the lack of RV measurements prevent us from calculating the orbit, the low Galactic height ($z\sim -1.1$~kpc) and the very small vertical proper motion directed toward the Galactic plane ($\mu_\mathrm{b}$=+0.223~mas~yr$^{-1}$) suggest a possible association with the old Galactic disk. 
The combination of old age, high metallicity, and disk orbit closely resembles 47\,Tuc and other three GCs (namely M107, NGC\,6362, and E3), whose polar paths cross in two very well-defined points on the sky \citep{2015A&A...581A..13D}. Nevertheless, we have checked that the polar path of Garro01 has no similarity with theirs, and it does not pass near these intersections. Therefore, the association of this new GC with the possible Sagittarius stream proposed by \citet{2015A&A...581A..13D} should therefore be excluded.\\

\section{Conclusions}
We report the discovery of the GC Garro01  in the VVVX near-IR images. This is a low-luminosity GC, located at a low latitude in the MW disk, with large reddening and high foreground stellar density. We complement our VVVX dataset with 2MASS near-IR photometry as well as Gaia optical photometry and PMs.

The optical and near-IR photometry allowed us to construct CMDs, 
in order to estimate principal GC parameters, such as age, metallicity, distance, absorption and reddening.
Most field stars are less reddened and therefore located in the foreground disk.

The optical and near-IR CMDs show that this new GC is more metal-poor and older than the canonical old metal-rich open cluster NGC 6791 ($D=4.1$ kpc, $t=9$ Gyr, $[Fe/H] =0.3$ dex, \citealt{Boesgaard_2009}). On the other hand, we find close similarity with the physical parameters of 47 Tuc ($D=4.5$ kpc, $t=11.8$ Gyr, $[Fe/H]=-0.7$ dex, \citealt{10.1093/mnras/stx378}), arguing for an old GC at $D=15.5$ kpc. However, this cluster is intrinsically 4.16 mag fainter than 47 Tuc, implying that its total mass is about $M \approx 10^4$ $M_{\odot}$ (if 47 Tuc has $M=7\times10^5$ $M_{\odot}$, \citealt{Marks_and_Kroupa_2010}), on the low mass end of the Galactic GCs \citep{2019MNRAS.482.5138B}.
We also note the presence of two candidate RR Lyrae stars within 10 arcmin from the cluster centre. One of them has a distance consistent with the measured RC distance for the cluster. Its PMs, however, suggest that both RR Lyrae are not cluster members.  

The Galactic GC census is incomplete at low latitudes \citep{Ivanov_2005, Minniti_2017}, and the discovery of Garro01 suggests that there are more low-luminosity GCs may still be uncovered at low Galactic latitudes. 

Another interesting implication is that the location, distance and metallicity of this GC match those of the MRi structure, a potential accretion event recently identified in the Galactic plane. The possibility of a physical association with this structure must be confirmed with follow-up spectroscopy to measure radial velocities and detailed chemical abundances.

\begin{acknowledgements}
ERG acknowledges support from an UNAB PhD scholarship.
J.A.-G. acknowledges support from Fondecyt Regular 1201490 and from ANID, Millennium Science Initiative ICN12\_009, awarded to the Millennium Institute of Astrophysics (MAS).
DM and MG are supported by Proyecto FONDECYT Regular No. 1170121.
DM is also supported by the BASAL Center for Astrophysics and Associated Technologies (CATA) through grant AFB 170002. 
We gratefully acknowledge the use of data from the ESO Public Survey program IDs 179.B-2002 and 198.B-2004 taken with the VISTA telescope and data products from the Cambridge Astronomical Survey Unit. PWL acknowledges support by STFC Consolidated Grant ST/R00905/1. SRA acknowledges support from the FONDECYT Iniciacion project 11171025, the FONDECYT Regular project 1201490, and the CONICYT + PAI ``Concurso Nacional Insercion de Capital Humano Avanzado en la Academia 2017'' project PAI 79170089. R.K.S. acknowledges support from CNPq/Brazil through project 305902/2019-9.
\end{acknowledgements}

\bibliographystyle{aa.bst}
\bibliography{bibliopaper}

\onecolumn

\begin{table}[h]
\centering 
\caption{Final GC physical parameters.}
\begin{tabular}{lc}
\hline\hline
\textbf{Parameter} & \textbf{Value} \\
\hline
RA (J2000) & 14:09:00.0  \\
DEC (J2000) & -65:37:12  \\
Latitude & $ 310.828^{\circ} $\\
Longitude  & $ -3.944^{\circ} $\\
$\mu_{\alpha^\ast}$ [mas yr$^{-1}$]& $-4.68\pm 0.47$ \\
$\mu_{\delta}$ [mas yr$^{-1}$]& $-1.35 \pm 0.45 $ \\
$A_{K_s}$ [mag]& $ 0.15\pm 0.01$ \\
$E(J-K_{s})$ [mag]& $ 0.30\pm 0.03$  \\
$(m-M)_{0}$ [mag] & $ 15.93\pm0.03 $\\
D [kpc] &$ 15.5\pm 1.0 $\\
$R_G$ [kpc] &$ 11.2\pm 0.2 $\\
Height $z$ [kpc] &$1.0$\\
$M_{K_s}$ [mag]& $-7.76\pm0.5$ \\
$M_{V}$ [mag] & $-5.26\pm1.0$ \\
$[Fe/H]$ [dex]& $ -0.7\pm0.2$ \\
Age [Gyr]&  $11.0\pm 1.0$  \\
$r_c$ [arcmin] & $2.1\pm 1.5$ ($4.6\pm 3.1$ pc)\\
$r_t$ [arcmin] & $6.5^{+11}_{-1.9}$ ($15^{+25}_{-4}$ pc)\\ 
\hline\hline
\end{tabular}
\label{table1}
\end{table}

\end{document}